\documentclass[9pt,twocolumn,twoside]{osajnl}

\journal{ol} 

\setboolean{shortarticle}{true}

\usepackage{lineno}

\title{Cold-atom shaping with MEMS scanning mirrors}

\author[1,*]{Alan Bregazzi}
\author[2]{Paul Janin}
\author[1]{Sean Dyer}
\author[1]{James. P. McGilligan}
\author[1]{Oliver Burrow}
\author[1]{Erling Riis}
\author[2]{Deepak Uttamchandani}
\author[2]{Ralf Bauer}
\author[1]{Paul. F Griffin}

\affil[1]{SUPA and Department of Physics, University of Strathclyde, G4 0NG, Scotland, UK}
\affil[2]{Department of Electronic and Electrical Engineering, University of Strathclyde, Glasgow, G1 1XW, Scotland, UK}

\affil[*]{Corresponding author: alan.bregazzi@strath.ac.uk}

\begin{abstract}
We demonstrate the integration of micro-electro-mechanical-systems (MEMS) scanning mirrors as active elements for the local optical pumping of ultra-cold atoms in a magneto-optical trap. A pair of MEMS mirrors steer a focused resonant beam through a cloud of trapped atoms shelved in the \textit{F}=1 ground-state of  \textsuperscript{87}Rb for spatially-selective fluorescence of the atom cloud. Two-dimensional control is demonstrated by forming geometrical patterns along the imaging axis of the cold atom ensemble. Such control of the atomic ensemble with a microfabricated mirror pair could find applications in single atom selection, local optical pumping and arbitrary cloud shaping. This approach has significant potential for miniaturisation and in creating portable control systems for quantum optic experiments.
\end{abstract}

\setboolean{displaycopyright}{true}

\begin{document}

\maketitle

\section{Introduction}
Much effort over the past decade had been directed towards the leveraging of well-established silicon processing techniques for the miniaturisation of laser cooling platforms, with recent research addressing compact vacuum cells \cite{McGilligan2020,boudot2020,Bregazzi2021}, optical components \cite{Nshii2013,chauhan2019photonic,Sitaram2020,Zhu2020,Mcgehee2021} and magnetic trapping on `atom chips' \cite{reichel1999atomic,folman2008microscopic}. One such technology driving the minaturization of cold atom sensors is the grating magneto-optical trap (GMOT) \cite{Nshii2013}, enabling use of a single beam incident on a microfabricated diffraction grating where the diffracted orders of light satisfy conditions for three-dimensional laser cooling in a highly compact manner. When paired with other compact technologies the possibility for truly portable cold atomic devices such as clocks and interferometers and cold atom quantum memories operating outside the lab environment is now beginning to emerge \cite{muller2009,Sorrentino2010,Chanelire2005}.

One growing field of interest within the miniaturisation of these platforms is that of optical manipulation of cold atom ensembles. Fine control of atom distributions by optical manipulation is well established and used extensively throughout quantum optic experiments from simple dipole traps \cite{grimm2000optical} and optical tweezers \cite{Barredo2016}, whose position is typically controlled by orthogonal acousto-optic modulator (AOM) pairs, to more complex arbitrary atom arrays formed by spatial light modulators (SLMs) or digital micro-mirror devices (DMDs) \cite{Nogrette2014,stuart2018}. These techniques also allow for shaping of time averaged potentials, for example by scanning orthogonal AOM pairs \cite{henderson2009,bell2016} while simultaneously providing a means to optically shutter the incident light.

When transitioning towards more portable cold atom devices where size, weight and power (SWaP) constraints must be more closely considered, many of these current solutions to optical manipulation and beam shuttering are non-ideal. 
Previous research has demonstrated the suitability of individual scanning MEMS micro-mirrors as an alternative for the addressing of single, neutral atom or ion sites where two orthogonally scanning MEMS mirrors are used to provide two-dimensional (2D) control of beam position \cite{Knoernschild2010,Crain2014,Wang2020}. Here we demonstrate a proof of principle investigation of 2D time averaged painted light potentials by selectively painting hyperfine repumping light onto a cold atom ensemble using two silicon-on-insulator MEMS micro-mirrors. By independently controlling each mirror we locally excite and image fluorescence within the atom cloud in line, square and circular geometries with scan frequencies close to 90~kHz. This technique is highly scalable, low SWaP, and offers the opportunity for simultaneous control of multiple wavelengths of light; typically more difficult to achieve with AOM control \cite{Kim2008}.

\section{Experimental set-up}

\begin{figure}[tbp]
    \centering
    \fbox{\includegraphics[width=0.9\linewidth]{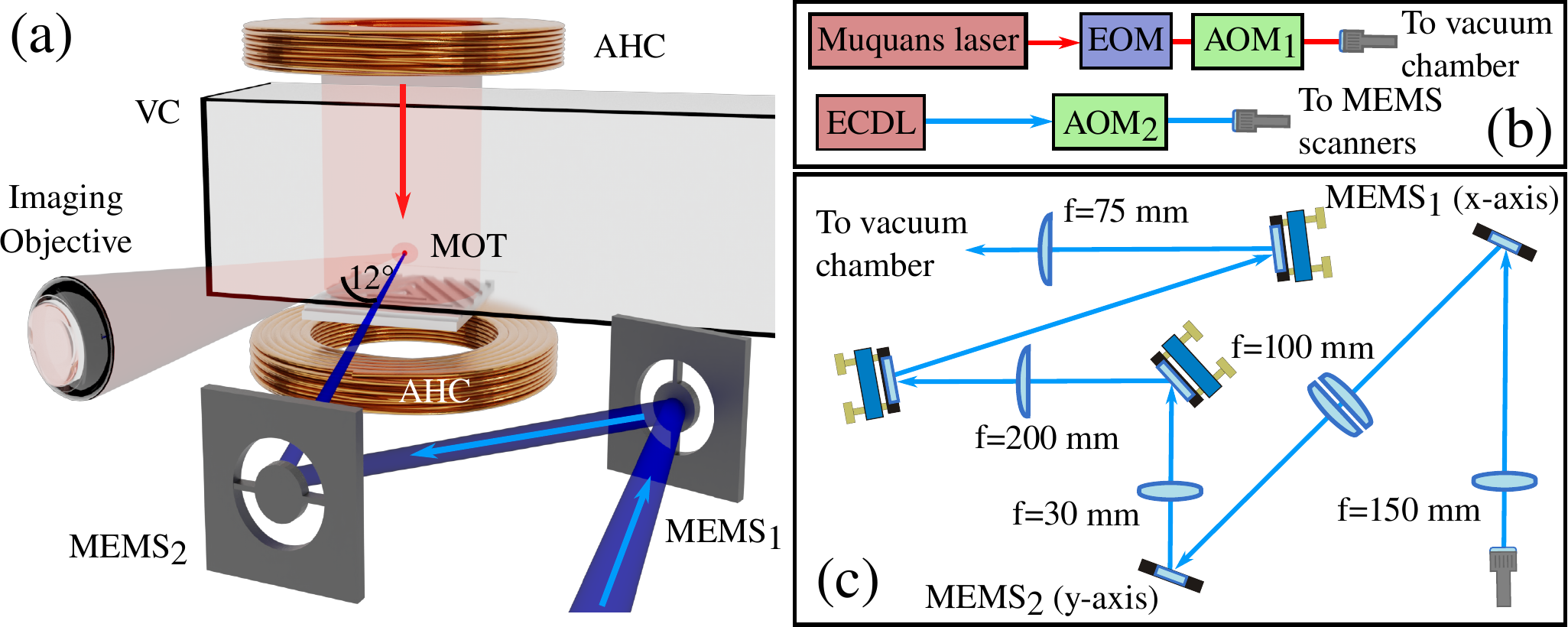}}
    \caption{(a) Simplified experimental set-up. Single trap beam (red) is aligned onto grating chip mounted externally to vacuum cell (VC) also shown are the anti-Helmholtz trapping coils (AHC). MEMS addressing beam (blue) is shown aligned into the MOT region with two orthogonally scanning MEMS mirrors. The imaging axis is shown rotated by $\approx$12\textdegree~from the MEMS addressing axis. (b) Schematic of main trapping laser (red) and MEMS addressing beam (blue) set-ups. EOM and AOM\textsubscript{1} are shown externally to the laser for clarity. (c) Overview of full MEMS scanning system, showing the re-pump beam preparation and alignment using two MEMS scanners.}
    \label{fig:exp set-up}
\end{figure}

As a basis for the proof-of-principle demonstration we use a grating magneto-optical trap formed by a 2~cm$\times$2~cm three zone "tri" grating \cite{Nshii2013}. A commercial laser system (Muquans ILS) outputting fibre coupled light at 780~nm is used throughout as a source of trapping, imaging light and main re-pump light. This allows for laser frequency tuning of $\approx$1~GHz over the whole \textsuperscript{87}Rb D\textsubscript{2} line with an internal acousto-optic modulator (AOM) for power control and an electro-optic modulator (EOM) to provide re-pump light during the trapping process. From the fibre we expand the beam to a $1/e^2$ radius of $\approx$2~cm, circularly polarize it with a quarter wave plate and align onto the grating. A simplified schematic of the experimental set-up is shown in Fig.\ref{fig:exp set-up}~(a), highlighting the optical axis for cooling, MEMS addressing and imaging.

The experimental sequence is initiated by turning on a pair of anti-Helmholtz trapping coils producing a gradient field of $\approx$15~G/cm. The laser frequency is set to be $\Delta=$2$\Gamma$ red detuned ($\Gamma/2\pi=$6.07~MHz) from the \textit{F}=2$\rightarrow$ \textit{F'}=3 cycling transition, while re-pump is provided by the EOM, modulated at 6.57~GHz to produce sidebands on the carrier light with 5\% of the total power. We load $\approx$2$\times10^6$ atoms into the GMOT in 2~s before beginning a 10~ms initial red molasses cooling stage. During this initial cooling, the trapping coils are turned off and the carrier detuning is linearly ramped to -5$\Gamma$ while simultaneously decreasing the beam intensity from 12$I_{\rm{sat}}$ to 3$I_{\rm{sat}}$, where the average $I_{\rm{sat}}$=3.5~mW/cm\textsuperscript{2} over all polarisations is used \cite{mcgilligan15}. We then perform a second shorter grey molasses cooling stage \cite{Barker2022} where the laser frequency is jumped to 5$\Gamma$ blue detuned from the \textit{F}=2$\rightarrow$\textit{F'}=2 transition with the EOM frequency changed to 6.83~GHz to concurrently tune the re-pump frequency. The laser intensity is also raised to 12$I_{\rm{sat}}$ and linearly scanned back down to 3$I_{\rm{sat}}$ in 3~ms. This second molasses stage was utilised due to the higher atomic densities typical of grey molasses \cite{Rosi2_grey_mol}, giving higher signal to noise ratio in the resulting MEMS re-pump addressing images. In addition to this, grey molasses efficiently pumps the atoms into the dark \textit{F}=1 ground-state \cite{Rosi2_grey_mol}, helping to prepare them for the MEMS addressing stage. Once the atom cloud has been prepared in the \textit{F}=1 ground-state the main re-pump light derived from the EOM is turned off with an extinction of 65~dB to ensure that the only re-pump light incident on the atoms originates from the MEMS addressing. We then wait 1~ms with the cooling beam on to ensure as many atoms as possible are shelved in the \textit{F}=1 ground-state before turning the MEMS addressing beam on. 

The MEMS addressing beam is derived from a home built extended cavity diode laser (ECDL) \cite{Arnold1998} locked via saturated absorption spectroscopy. The power level of this light is controlled using a single pass AOM that can be rapidly turned on and off. This light, now resonant with the D\textsubscript{2} \textit{F}=1$\rightarrow$\textit{F'}=2 transition, is then fibre coupled and passed to the MEMS mirror scanning system.

\begin{figure}[tbp]
    \centering
    \fbox{\includegraphics[width=0.9\linewidth]{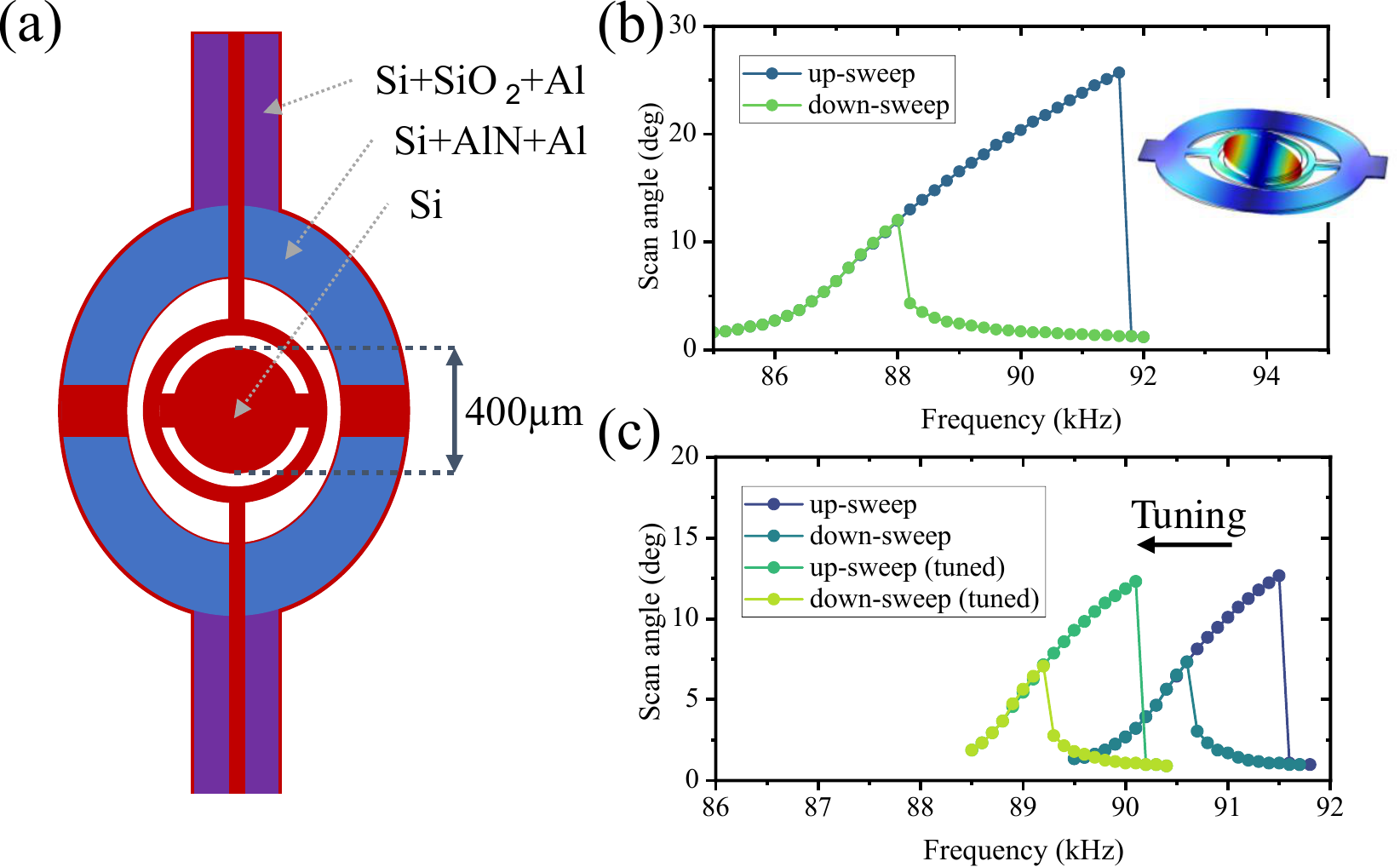}}
    \caption{MEMS mirror characteristics. (a) Schematic of the used 400 $\mu$m diameter mirror. (b) Frequency response for MEMS\textsubscript{1}. (c) Frequency response for tuning of MEMS\textsubscript{2}.}
    \label{fig:mems}
\end{figure}

The scanning system is based on a pair of resonant MEMS scanners to steer the MEMS addressing beam onto the cold atom cloud. The fiber-coupled laser beam is collimated into free space with a waist of 1~mm and focused to a point roughly 30~mm before the surface of the first MEMS scanner (MEMS\textsubscript{1}) by a 150~mm focal length achromatic lens. The beam reflected from MEMS\textsubscript{1} is imaged onto the second MEMS (MEMS\textsubscript{2}) by a matched achromatic doublet pair with 100~mm focal length, ensuring that beam steering deflections originate from the same point on the optical axis. The MEMS are positioned to align their scan axes orthogonally. The scanner image is then relayed through a 30~mm focal length achromat followed by a 200~mm plano-convex lens, such that the final scanner image is placed in the object plane of the final focusing lens; creating a telecentric illumination. Furthermore, placing the scanner out of focus of the initial lens allows collimation of the beam before the final imaging lens, and focusing at the target, such that scanline focusing is decoupled from the beam focusing.

The scanners themselves are resonant piezoelectric MEMS with 400~$\mu$m  mirror aperture fabricated using a commercial multiuser process (MEMSCAP PiezoMUMPS). The scanner design is shown in Fig.\ref{fig:mems}~(a). A dual frame allows for placement of four aluminium nitride piezoelectric actuators on the outer frame, coupling and magnifying its out of plane rotation motion to the mirror surface (see inset Fig.\ref{fig:mems}~(b)). In the following, both MEMS are actuated around the fundamental tilt resonance mode of the mechanical structure, rotating the mirror along the inner frame suspension beams connecting the mirror surface to the inner frame. The corresponding resonance frequency is around 90~kHz with a maximum optical scan angle of 25\textdegree~ and 12\textdegree~ using an offset sine-wave actuation with 40~V$_{pp}$ and 20~V$_{DC}$ offset (see Fig.\ref{fig:mems}~(b) and (c)). In both cases a hysteresis behaviour of the frequency response is apparent, with nonlinearities leading to a higher movement scan angle being reachable when approaching the resonance peak from lower frequencies. For MEMS\textsubscript{1} this leads to the larger angular response in the frequency range between 88 and 92~kHz to only be reachable when starting the movement below 88~kHz, while for MEMS\textsubscript{2} the same behaviour is present in the frequency range between 90.5 and 91.5~kHz.

In order to compensate for discrepancies in resonance frequencies between the two scanners resulting from variations and tolerances during the fabrication process, MEMS\textsubscript{2} can be tuned by localized heating of the device through three on-chip thermal actuators. This concept is described in detail in \cite{Janin2022}. Through thermal tuning, the resonance frequency can be reduced by around 1.5~kHz, from 91.5 to 90~kHz, making actuation at 90~kHz possible with sufficient optical scan amplitude for synchronised scanning purposes.

Imaging of the atom cloud was preformed by standard fluorescence detection on a CCD camera with the imaging axis offset by 12\textdegree ~from the addressing axis due to optical access.

\begin{figure*}[tbp]
    \centering
    \fbox{\includegraphics[width=0.9\linewidth]{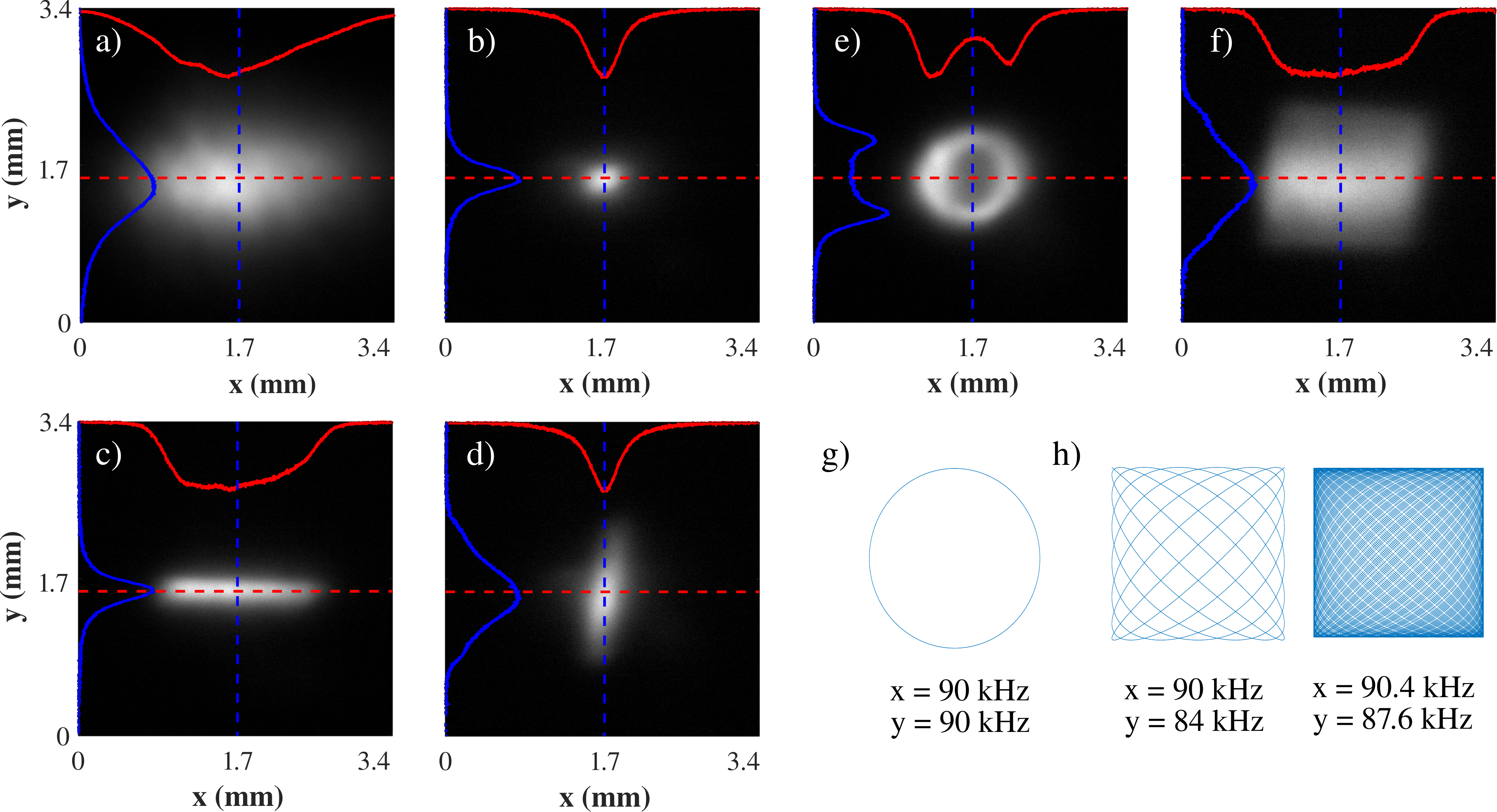}}
    \caption{Images of atomic fluorescence for $\approx2\times 10^6$ atoms loaded into the standard GMOT (a), region of atom cloud illuminated by MEMS addressing beam with no MEMS scanning (b), horizontal MEMS mirror scan (c), and vertical MEMS mirror scan (d), circular scan driving both MEMS with the same frequency and pi/2 phase shift (e), square scan using both MEMS with Lissajous scanning (f). All images are individually normalised. The traces in (g) and (h) show the 2D scan profiles for the atomic fluorescence profiles in (e) and (f) respectively, with (h) highlighting a general Lissajous scan profile for illustration next to the used profile.}
    \label{fig:images}
\end{figure*}

\section{Results}

An image of the whole atom cloud after optical molasses is shown in Fig.\ref{fig:images}~(a). This image was taken with the main experimental re-pump light, derived from the EOM, incident on the atoms and shows the extent of the atom cloud to be roughly 3~mm in the horizontal axis and 1.6~mm in the vertical axis. Fig.\ref{fig:images}~(b) shows an image of the atom cloud after extinguishing the main re-pump light and turning the MEMS addressing beam on with no MEMS mirror scanning present. When a Gaussian distribution is fitted to the vertical axis of the image in Fig.\ref{fig:images}~(b) a waist of 300~$\mu$m is calculated. This differs to measurements of the MEMS addressing beam giving a beam waist of 40~$\mu$m. The apparent broadening is largely attributed to the relatively long exposure time of the fluorescence images of 1~ms and the finite temperature of the atom cloud of 50~$\mu$K, resulting in the motion of atoms within this time window.

By actuating each MEMS mirror individually we can create single line scans in the horizontal and vertical axis, as shown in Fig.\ref{fig:images}~(c) and (d) respectively. These narrow line potentials are again broadened with respect to the incident beam, giving waists of 260~$\mu$m for the horizontal scan and 380~$\mu$m for the vertical scan, with a scan length of 2~mm and 1.4~mm respectively. The additional broadening of the vertically scanned beam is likely due to the imaging axis being rotated with respect to the incident beam. Line profiles showing the intensity profiles through the center of the cloud are shown by the red and blue overlays in Fig.\ref{fig:images}~(a)-(f). Local fluorescence within the atom cloud results from a combination of the local atomic density and the time averaged beam intensity at each location along the scan axis. Due to the rotation of the mirrors around their fundamental tipping mode, the addressing beam spends more time interacting with atoms at the extremes of the scan. By contrast however there is a lower atomic density at the edges of the scans due to the MOT's Gaussian profile. The combination of these effects in the ideal case leads to a flattened intensity profile around the centre of the MOT with Gaussian edges originating from the Gaussian addressing beam as shown in Fig.\ref{fig:images}~(c). When scanning vertically as in Fig.\ref{fig:images}~(d) the intensity profile can be seen to diverge significantly from the ideal square case. This discrepancy between the horizontal and vertical scans is due to the characteristic "pancake" shape of GMOTs, where slightly larger axial trapping forces compared to radial trapping cause the MOT to be compressed in the vertical axis \cite{mcgilligan15}. When the scan length is of the same order as the atom cloud's dimension in that axis the larger variation in atomic density across the scan length causes this deviation in the scan profiles.

Next the mirrors were simultaneously driven around their respective resonance frequencies (87.6~kHz for MEMS\textsubscript{1} and 90.4~kHz for MEMS\textsubscript{2}), producing a time averaged square light potential within the cloud, shown in Fig.\ref{fig:images}~(f). This results from the Lissajous scan pattern with $1.97 \times10^{5}$ cross-over points per axis, generated by the precise frequency spacing between both base resonances. This concept is illustrated in Fig.\ref{fig:images}~(h) where simulations of the resulting scan profile are shown for two different scan frequencies. Line profiles through each axis of this image once again show a good approximation to a square profile in the horizontal axis with the vertical axis deviating significantly from the ideal square potential.


 A circular scan, shown in Fig.\ref{fig:images}~(e), is generated when the resonance frequencies of both MEMS mirrors are identical, with a 90\textdegree~ phase shift between their movements. To achieve this the frequency tuning actuation of MEMS\textsubscript{2} was utilised, with a 7.5~V$_{DC}$ tuning voltage applied to one of the thermal tuning actuators and 10~V$_{DC}$ to the other two, while MEMS\textsubscript{1} was left at its original resonance. This led to a matched resonance frequency and therefore circular scan frequency of 90~kHz, with the scanning beams creating a trace for a ring potential. One requirement for optical ring potentials within BECs for interferometry is that the light potential is smooth across the whole profile of the ring. While Fig.\ref{fig:images}~(c) shows a clear imbalance in intensity of the left hand side of the ring compared to the right, this effect is primarily due to the 12\textdegree~rotation in the imaging axis with respect to the MEMS addressing beam. This rotation means the three-dimensional hollow cylinder of bright atoms is projected into the imaging plane at an angle. One side of the 2D image is therefore integrated over a higher density of bright atoms than the other side, leading to one side of the image appearing much brighter than the other. Another contributor to this intensity imbalance is local variations in atomic density, visible in Fig.\ref{fig:images}~(a). Previous demonstrations of ring potentials have accounted for these local variations in density by modulating the light intensity as it is scanned with orthogonal AOMs \cite{bell2016}. Implementing some feedback on light intensity the 2D MEMS mirror scanner would also serve to create smoother potentials for use in portable quantum optic experiments, with required modulation frequencies of 90~kHz within easy reach of direct diode modulation schemes.

\section{Conclusions}

In conclusion, we have demonstrated a proof of principle experiment for shaping time averaged light potentials in cold atom ensembles using MEMS mirrors. By independently driving two orthogonally scanning MEMS mirrors with resonance frequencies above 90~kHz we are able to form simple line, square and circle profiles of bright fluorescing atoms within a larger dark atom cloud. This technique could be extended to dynamic beam shuttering and further demonstrates the applicability of MEMS microfabrication techniques for replacing high SWaP lab based equipment, building on previous research towards the miniaturisation of cold atom experiments for the realisation of truly portable cold atom sensors. The wide bandwidth of available reflection coatings on the MEMS mirrors make this technique ideal for simultaneous control of multiple wavelengths of light, while more dynamic control of the scan amplitude and phase would make it possible to form more complex geometries opening the technique to more applications such as the storing of images for quantum information processing \cite{Shuker2008}.

\begin{backmatter}

\bmsection{Acknowledgments} The authors acknowledge funding from the Engineering and Physical Science Research Council (EP/T001046/1, and EP/S032606/1) as well as from the UK Royal Academy of Engineering Research Fellowship scheme (RF1516/15/8 for R. B., and RF/202021/20/343 for J. P. M). A. B. was supported by a PhD studentship from the Defence Science and Technology Laboratory (Dstl) and P. J. was supported by a studentship from the National Physical Laboratory (NPL).

\bmsection{Disclosures} The authors declare no conflicts of interest.

\bmsection{Data availability} Data underlying the results presented in this paper are available at https://doi.org/10.15129/4f23fe3b-e52c-43fd-8206-edc5806b2e50.

\end{backmatter}

\bibliography{Ref}


\end{document}